# Rapid Integration and Calibration of New Sensors Using the Berkeley Aachen Robotics Toolkit (BART)

Jan O. Biermeyer[†]   Todd R. Templeton[†]   Christian Berger[‡]
Humberto Gonzalez[†]   Nikhil Naikal[†]   Bernhard Rumpe[‡]   S. Shankar Sastry[†]

[†]Department of Electrical Engineering and Computer Sciences, University of California, Berkeley, CA 94720, U.S.A.
phone: +1 (510) 643-9783, fax: +1 (510) 643-2356, email: {janb, ttemplet, hgonzale, nnaikal, sastry}@eecs.berkeley.edu.

[‡]Software Engineering, RWTH Aachen, 52074 Aachen, Germany.
phone: +49 (241) 80-21301, fax: +49 (241) 80-22218, web: http://se-rwth.de/.

## Abstract

After the three DARPA Grand Challenge contests many groups around the world have continued to actively research and work toward an autonomous vehicle capable of accomplishing a mission in a given context (e.g. desert, city) while following a set of prescribed rules, but none has been completely successful in uncontrolled environments, a task that many people trivially fulfill every day. We believe that, together with improving the sensors used in cars and the artificial intelligence algorithms used to process the information, the community should focus on the systems engineering aspects of the problem, i.e. the limitations of the car (in terms of space, power, or heat dissipation) and the limitations of the software development cycle. This paper explores these issues and our experiences overcoming them.

# 1 Introduction

After the 2007 DARPA Urban Challenge [1], we summarized our ideas in a position paper [2] at AAET 2008. We greatly benefited from the subsequent discussions and interactions at AAET 2008 and 2009. With this paper, we return to AAET 2010 to present our progress, and to invite feedback on our achievements and our proposed future research directions from both academia and industry.

We focus on the following accomplishments since our first AAET paper: *(i)* We have fully actuated and equipped a 2008 Ford Escape Hybrid XGV (Figure 1), which is capable of remote control, assisted driving, and fully autonomous operation; we describe it in detail in Section 2. *(ii)* We have developed the Berkeley Aachen Robotics Toolkit (BART), which is comprised of Berkeley's Intelligent Robotics Toolkit (IRT) and Aachen's Hesperia Software Environment. We describe BART in Section 3, IRT in Section 3.1, and Hesperia in Section 3.2. We hope to soon release BART to the general public under a three-clause BSD license, and we actively support and welcome its use in both academia and industry, including as a royalty-free toolkit for teaching purposes; we also welcome the integration of new sensor drivers and high-level algorithms.

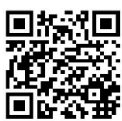

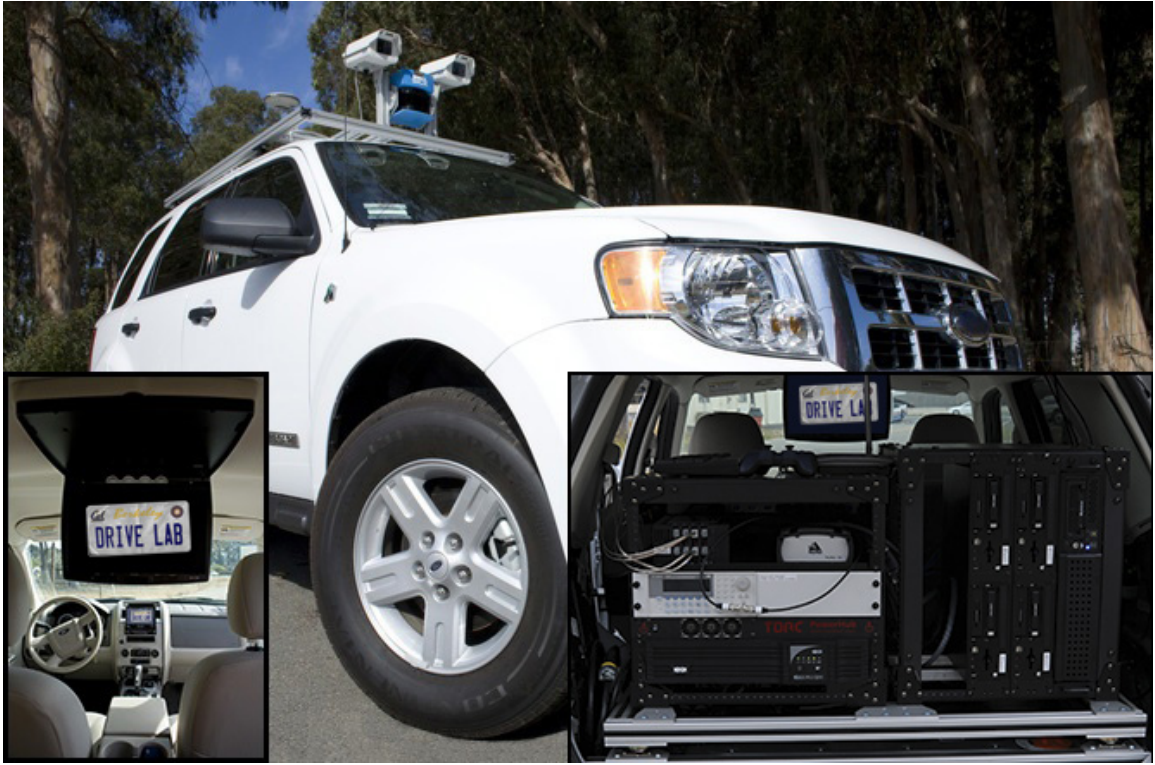

Figure 1: Berkeley DRIVE LAB test platform: 2008 Ford Escape Hybrid XGV

To demonstrate the ease-of-use of our hardware and software platform, we show in Section 4 the rapid integration of an automotive sensor, the Hella IDIS laser scanner, as well as our approach to calibrating it with respect to a camera system.

## 2 Hardware Platform

In this section, we describe our vehicle along with other potential candidates and actuation strategies. We also discuss our computing infrastructure and our sensors.

### 2.1 Vehicle Selection and Actuation

From our experience using an experimental vehicle for the first two DARPA Grand Challenges [3, 4] and a self-actuated vehicle at the DARPA Urban Challenge [5], it was clear that a car actuated by an OEM, or a specialized kit with support from a third-party supplier, would be the best choices for minimizing actuation time and maximizing reliability and usable test time.

Our main criteria for selecting a vehicle were *(i)* energy efficiency, *(ii)* ease of actuation, *(iii)* unobstructed manual operation, *(iv)* agility, *(v)* comfortable seating for four adults, *(vi)* spacious trunk for computing equipment, *(vii)* low vibrations when engine on, and *(viii)* California street legal.

Examining the success of DARPA Urban Challenge teams without previous autonomous car infrastructure, the Volkswagen Passat and Tiguan were obvious choices because of VW's extensive experience actuating vehicles, the fuel efficiency of the turbo diesel injection (TDI) engine, the optimization of the "Darmstadt" steering column for computer-controlled actuation, the available CAN gateway and an electrical outlet in the trunk, and the fast vehicle delivery (only a few weeks). The main drawbacks of a Volkswagen were its relatively high price (partially due to the weak dollar) and Berkeley regulations favoring hybrid vehicles.

With the US military as a major potential sponsor, a military Humvee (Hummer) was another obvious choice. Other advantages of the Humvee were its performance in rugged terrain and spacious seating for in-car debugging. For automation, we considered the Kairos strap-on autonomy kit from Kairos Autonomy, a Utah-based company; the autonomy kit could be adapted to virtually any vehicle, but was mainly developed on a Humvee. Drawbacks of the Humvee included its large turning radius, which would complicate navigation in tight urban spaces; vibration that would make high-accuracy sensing more difficult; and the dubious street legality of the strap-on autonomy kit, as well as its obstruction of manual operation.

The Ford Escape Hybrid was one of the first X-by-Wire cars on the US market. For actuation, the obvious choice was TORC Technologies' ByWire XGV solution, which used the experience of the DARPA Urban Challenge team Victor Tango to manually actuate the brake and control all other functions via the ECU. Advantages of this vehicle and autonomy solution included the fact that the actuation would not obstruct manual operation of the vehicle, that its ground clearance would allow it to handle occasional grass fields and washed-out dirt roads (if not true off-road use), and that electricity could be drawn from the hybrid battery in the trunk. The major downside was that trunk space would be somewhat limited because of the large hybrid battery; however, we would would still be able to fit two shock-mounted computer racks in the trunk.

After weighing the alternatives, we chose the Ford Escape Hybrid / ByWire XGV solution. To compensate for the fact that the outer hull was made of thin sheet metal to save weight, we contracted with a local automotive shop to reinforce roof attachment points for our sensor mounting rack and to install a bull bar on the front to minimize the potential for deformation during minor obstacle contact during testing. One problem to which we have not found an adequate solution is that the vehicle cannot be used for some types of driver assistance, since the steering wheel is actuated using the power steering motor, which has the potential to break a human arm; we are currently experimenting with top-of-the-line game controllers, such as the Logitech G25 Racing Wheel, as an alternative for prototyping such systems.

## 2.2 Computing and Electrical System

One of our first steps was to install two Star Case computer racks (20"x19" 8RU) in the trunk, on shock-absorbing elastomeric isolation mounts. We dedicated one rack to computers, which we mounted vertically to minimize the effect of any shock on the hard drives. We dedicated the other rack to auxiliary devices such as power-supply equipment and a signal generator (see Figure 1, bottom right inset). We contracted with TORC Technologies to install the power-supply equipment: a UPS that we could plug into the wall when the car was stationary, and a DC-DC converter to transform the 330V hybrid battery into an auxiliary battery for the UPS (which had its buzzer disabled) for normal operation. The installed power system was rated for 2500 Watts, which has proven to be more than adequate so far. In order to allow all of our equipment to run on either car or wall power, we decided to power everything through the UPS, instead of using DC power directly from the DC-DC converter for our DC sensors.

Our main criteria for our on-board computers were *(i)* high computing capacity, *(ii)* low energy consumption, *(iii)* small form factor, *(iv)* high reliability, *(v)* flash drives for the operating system (more robust to vibration), and *(vi)* easily-removable hard drives for collecting sensor data.

Because of our strict requirements, we decided to build our four base computers ourselves. We built them around Intel Q6600 Core 2 Quad chips on Mini-ITX boards, mounted into ITS-2814 Half-2U Mini-ITX cases from IdotPC International (which can be connected in pairs to form 2U assemblies). The boards provide dual Gigabit Ethernet and can hold 4 GB of single-sided RAM, although at the time we were only able to find 1 GB single-sided modules. The operating system, Debian Linux, is stored on Compact Flash (CF) cards for reliability; the CF cards are easy accessible from the outside of the case, for off-line copying as well as for fast switching between operating systems or different versions of the entire software system.

All data is logged on standard 2.5" notebook SATA hard disks in Vantec EZ-Swap EX enclosures. The EZ-Swap bays are connected to the on-board computers via SATA, and the drive inserts also fit into bays in off-line desktop computers or can be connected to notebooks via USB. The disks are mounted vertically to minimize drive failure; however, because the operating system is stored on the CF cards, hard drive failure will not disrupt operation (other than logging).

The latest addition to our onboard computers is a 2U full-width computer with dual Intel 50W TDP LV Xeon L5420 (2.50 GHz) quad core processors and 8 GB of memory—a customized Titan-G2X / 200 from GamePC.com (future units will utilize low-power Xeon multi-cores with HyperThreading support for even better computational efficiency). This computer also features an EZ-Swap bay for a data-logging hard drive, but the flash drive for the OS and BART is an internal flash SATA disk instead of a CF card. We use this machine to experiment with parallel algorithms on the included NVidia Quadro FX 3800 GPU, which has 192 600 MHz processors and 1 GB of GDDR3 RAM, using a maximum of 108 Watts. Utilizing AccelerEyes' Jacket, we can prototype massive parallel algorithms on the car in MATLAB, or we can create more performance-optimized implementations in CUDA. We also use Intel's Thread Building Blocks (iTBB) for parallel computation on the multi-core CPU, for computations with a lower computation-to-storage ratio than can be optimized by using a GPU.

The passenger cabin is divided into two zones (see Figure 1, bottom left inset). We re-placed the stock stereo with a Xenarc MDT-X7000 stereo with a fold-out 7-inch in-dash touchscreen display and front USB access; a Logitech diNovo Edge wireless Bluetooth keyboard with built-in touchpad is stored in the seat pockets. This allows high-level control of the autonomy or assistance system by the front passengers. During development, it also allows the display of debugging info or vehicle state data, in addition to minor debugging. For the backseat passengers, a Xenarc 1530YR 15.3-inch high-contrast and -brightness display folds down from the roof and is easily readable even in bright light, unlike most notebook computers. All vehicle computers, except for the one connected to the front display, are connected to this display and a keyboard with integrated touchpad via an IOGear GCS78KIT 8-port KVM switch. Additionally, any notebook within the car and any computer within our lab can connect to any car computer via several Ethernet cables or our in-car wireless router, and our Netgear JGS524 24-port Gigabit Ethernet switch.

Throughout our test building, 802.11G wireless Internet access is provided by several D-Link DWL-3200AP access points in wireless distribution system (WDS) mode. By using D-Link ANT24-0700 omni-directional high-gain antennas on the vehicle's roof, as well as a custom outdoor antenna on the top of our building, wireless access is available virtually anywhere on the test track. Beyond streaming live video data, this connection can be used for remote control, online updates of BART or the Linux OS, or general Internet access. However, for security reasons, the connection can be quickly unplugged, to separate the wired core computers from any auxiliary wirelessly-connected computers.

## 2.3 Environment Perception

As outlined in [2], our main goal is to demonstrate intelligent cars via relatively cheap, commercial off-the-shelf sensors. SICK LMS laser scanners are a staple in most autonomous cars for environment perception, as are our Point Grey Flea2 visible light cameras (mounted in Pelco enclosures). Thermal IR "night vision" is provided by FLIR ThermoVision A320G cameras. Global position data and inertial measurements are provided by a NovAtel SPAN / ProPak with a Honeywell HG1700 IMU. Our newest sensor addition is a automotive laser scanner, the Hella IDIS, which we discuss further in Section 4. All sensors are mounted on MayTec rails, which enable both fixed mounts and rapid prototyping / experimentation.

For higher-accuracy comparison data, we use two sensors that are more expensive—we do not envision these sensors as part of any autonomy or driver assistance solution. We use a NovAtel GPS receiver and 900MHz wireless modem for differential GPS, as well as a top-of-the-line NovAtel SPAN (Synchronous Position, Attitude and Navigation) GNSS+INS with a tactical grade iMAR IMU-FSAS and magnetic wheel sensors, for more accurate vehicle state data. Similarly, we utilize a Velodyne HD LIDAR for high-resolution 360 degree environment perception.

One of the most important features of our sensor setup is that all data is accurately and con-sistently timestamped by embedded systems before it arrives at the primary vehicle computers. The FireWire cameras are implicitly timestamped by an external trigger from our Agilent 33210A signal generator. All other production sensors (this set does not include the Velodyne, which produces an extremely large volume of data) are connected to NetBurner microprocessor boards by either serial or CAN. These NetBurner boards are also connected to the camera trigger via a hardware interrupt pin, which they use to maintain an accurate and consistent time. All of the data that passes through the NetBurner boards is timestamped and then sent out to any subscribed vehicle computer using the Spread Toolkit [6]. Although all sensor data is timestamped before it reaches the vehicle computers, we also synchronize the vehicle computers using the Network Time Protocol (NTP). We are considering using a Meinberg GPS synchronized NTP server in the future for high-precision time synchronization, as well as real-time remote control capabilities.

For more details on our sensing and navigation algorithms, see [2].

# 3 Berkeley Aachen Robotics Toolkit (BART)

The Berkeley Aachen Robotics Toolkit (BART) is the fusion of Berkeley's Intelligent Robotics Toolkit (IRT) and Aachen's Hesperia Software Environment. IRT and Hesperia were initially developed separately, but since they complement each other they have joined forces to create a stronger toolkit.

## 3.1 Intelligent Robotics Toolkit

The Intelligent Robotics Toolkit (IRT) was originally based on a codebase for autonomous helicopter mapping and landing that had overgrown its flat filesystem layout. We needed to organize the codebase hierarchically by function, generalize the build system (a static Makefile) to other platforms (we had problems even with other Linux systems) and separate it into different files for different modules, and create a test infrastructure to ensure consistent code quality. The final result can be broken down into two pieces: *(i)* an advanced infrastructure for code organization, reuse, building, and testing; and *(ii)* a set of software modules for robotics and artificial intelligence that use this infrastructure.

### 3.1.1 Infrastructure

The IRT infrastructure is centered around a build system that uses Python and SCons [7] for cross-platform portability. The build system reads a file in each directory that contains declarative sections about the targets that can be built in that directory, including their (compile-time, run-time, required, and optional) dependencies, which can be anywhere in the repository. By convention, less-common third-party dependencies are included in the repository's `third_party` directory, and build rules for finding more-common third-party dependencies are included in the repository's `third_party/external` directory. At compile time, we can specify either a specific target to build or allow the build system to build all targets in the repository; the build system tells us what targets it is able to build, and the reason why it is unable to build some specific targets, e.g. a missing dependency. We can also specify a cross-compilation, e.g. to a specific NetBurner architecture.

We support C/C++, Python (including C/C++ Python extensions), and MATLAB (including C/C++ MATLAB extensions). Unit tests are built into the IRT infrastructure—we use Google's gtest C++ unit test library [8] and the built-in Python unittest package. At the end of every build, the build system creates a Python script that can be used to run all of the unit tests that were compiled, in addition to a shell script or batch file that can be used to set all necessary environment variables.

For cross-platform portability, we rely on our architecture compatibility layer (archcompat), which is a thin wrapper over the system's API to make it compatible with the standard Linux API. For communication, we use our chardevice library, which provides both cross-platform portability and the ability to seamlessly switch between different communication protocols, such as serial, TCP, UDP, and the Spread Toolkit [6]. For any remaining portability issues, we rely on our build system's ability to create configuration C/C++ header files or Python modules based on conditions such as whether certain external libraries are found; these configuration modules can be used for conditional compilation (C/C++) or run-time adaptation (Python).

To enable further modularity within the project, we use the concept of plugins. Our plugins system allows a subclass to register that it implements a specific interface, which allows its library to be dynamically loaded at runtime if the user requests its specific implementation, e.g. the UDP implementation of chardevice or the SICK implementation of the laser scanner driver. Our plugins library makes it easy for a new base class to support plugins using only a few simple macros.

### 3.1.2 Modules Overview

In this section we give a brief overview of our most important software modules:

**Communication**

| | |
|---|---|
| `cdterm` | terminal program that supports all chardevice protocols (GUI for chardevice) |
| `chardevice` | transparent and cross-platform communication via for serial, TCP, UDP, and the Spread Toolkit [6], with data logging and playback |

**Control**

| | |
|---|---|
| `pathplaner` | implementation of the model predictive controller (MPC) [2] |

**Math**

| | |
|---|---|
| `graphcluster` | graph segmentation—implementation of [9] |
| `image` | object-oriented, multi-type image library based on OpenCV [10] |
| `matrix` | object-oriented matrix library based on LAPACK [11], BLAS [12], and OpenCV [10] |
| `maxflow` | graph max flow—implementation of [13] |

**Sensor**

| | |
|---|---|
| `camera` | driver for FireWire cameras |
| `common` | common sensor driver functionality, including client/server |
| `ins` | driver for several INS devices |
| `ladar` | driver for several LADAR devices |
| `radar` | driver for several radar devices |

**Simulation**

| | |
|---|---|
| `imgrender` | renderer for sequences of synthetic images, using Blender [14], from texture and elevation images |

**Third Party**

| | |
|---|---|
| `gtest` | C++ unit-testing framework [8] |
| `external` | contains build rules for finding external libraries |
| `openjaus` | implementation [15] of the JAUS [16] component and communication architecture—used to communicate with the TORC XGV system |
| `scons` | cross-platform Python build tool—the base of our build system [7] |
| `sicktoolbox` | a single-threaded version of the Sick LIDAR Matlab/C++ Toolbox [17] |
| `spread` | Spread Toolkit for multicast communication [6] |
| `trio` | cross-platform stdio implementation [18] |

**Utilities**

| | |
|---|---|
| `archcompat` | architecture compatibility layer to ease portability between Linux, Unix, NetBurner, and Windows |
| `asyncproc` | asynchronous process with messaging, locking, and logging |
| `logger` | file logger |
| `messager` | inter-thread messaging |
| `plugins` | plugin system |
| `properties` | configuration file reader and generic properties interface |
| `repository` | base for repositories (such as frame repository in camera module), checkin/checkout interface with ability to write items to disk |

**Vehicle**

| | |
|---|---|
| `xgv` | joystick demo and JAUS-based XGV simulator |

**Vision**

| | |
|---|---|
| `checkerboarddetector` | checkerboard detector |
| `elevationmap` | modular elevation map used for terrain reconstruction |
| `rmfpp` | the Recursive Multi-Frame Planar Parallax (RMFPP) algorithm [19, 20] |

## 3.2 Hesperia Software Environment

The Hesperia Software Environment is a strictly object-oriented toolkit written in highly portable ANSI-C++ to support the development of distributed applications especially in real-time environments with embedded software. The main focus is on the virtualized software development for sensors- and actuators-based systems by providing an appropriate model of the system's context to be used for interactive and unattended system simulations. On the one hand, these simulations can be used interactively by the software developer to feed data into the system under development *(SUD)* for evaluation; on the other hand, these simulations can be used unattendedly to perform an evaluation automatically comparable to the well-known unit tests.

Hesperia was inspired by the experiences from the TU Braunschweig's contribution "Caroline" to the 2007 DARPA Urban Challenge. However, it was completely written from scratch extending and exchanging the concepts used for developing software for "Caroline."

### 3.2.1 Architectural Design of Hesperia

In Fig. 2, the core design of the Hesperia Software Environment is shown. It consists of two main libraries namely "libcore" and "libhesperia." The former library is the encapsulating library to the operating system or any hardware interfaces. Thus, it ensures platform independence and interoperability between heterogeneous systems by providing core point-to-point and broadcast communication concepts as well as thread-safe data storage and filter methods. Currently, this core library is available for Microsoft Windows XP, Microsoft Windows Vista, Microsoft Windows 7, openSUSE, Debian, Ubuntu, FreeBSD, and NetBSD.

The latter library, "libhesperia" is a further layer on top of the previous library. In this library, a domain specific language (DSL) for describing the system's context is integrated to model a system's context [21] for providing synthetic input data for various layers as described in the following. Moreover, a ready and easy to use concept for communication between distributed applications called "Client Conference" is provided. This concept allows a fast, extensible, scalable, and non-reactive communication for an unlimited number of participants.

On top of this library, the actual system is running. In general, sensors- and actuators-based autonomous systems can be divided into three major parts: A "Perception Layer" which perceives the system's surroundings by gathering and fusing raw sensor data, a "Decision Layer" which evaluates and interprets the abstract environmental model to derive an abstract action, and the final "Action Layer" which transforms the abstract action into necessary set values for the actuators and their controllers. To perform simulations to support the development of such a sensors- and actuators-based system, the aforementioned three layers must be closed altogether or separately into a loop; in the former case synthetic input data must be provided for the "Perception Layer" and the system's reaction must be evaluated at the "Action Layer" to generate new input data. In the latter case, layer-dependent input data must be provided. Therefore, the so-called "Virtualization Layer" is used to generate the necessary input data by evaluating the DSL which describes the system's context.

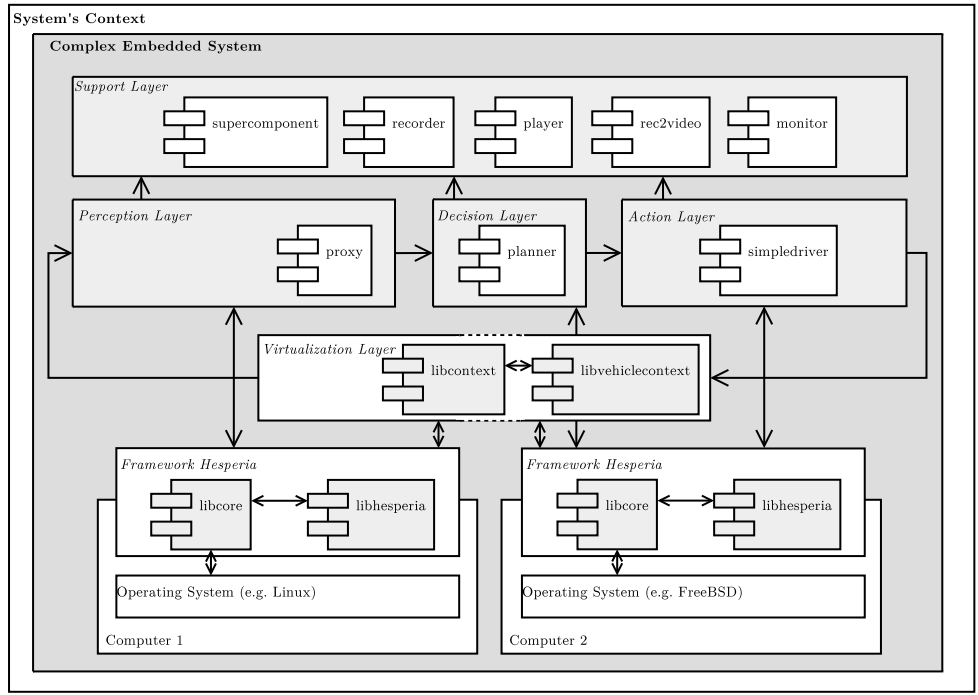

Figure 2: Overview of the architectural design of the software framework Hesperia: The framework consists of two main parts: "libcore" and "libhesperia." The former is a library for encapsulating and abstracting all interfaces to a specific operating system or hardware platform by providing convenient interfaces and wrapper classes; the latter is a library especially for supporting the development of sensors- and actuators-based autonomous systems by providing convenient concepts which reuse and further encapsulate the interfaces from "libcore." Moreover, "libhesperia" provides ready and easy to use, thread-safe communication concepts and data-storage.

For the running example of this paper, the library "libvehiclecontext" provides some models for the vehicle dynamics. For allowing unattended and automatable system simulations and evaluations, the library "libcontext" is used to abstract from the current real system clock and to control all running applications. Thus, comparable to unit tests, more complex system simulations can be described in a machine-processable manner to run and evaluate the SUD nightly or even more often by integrating into a continuous integration system.

For convenience, further applications are included in the Hesperia Software Environment. The application "supercomponent" is used to provide centralized and thus consistent configu-ration data using the "Dynamic Module Configuration Protocol" (DMCP) which is inspired by the well-known DHCP to configure remotely operating applications; furthermore, it supervises all running applications and tracks their life-cycle. The applications "recorder" and "player" are used to capture non-reactively all communication for later replay. The application "rec2video" is used to render a 3D video from a running system also using the aforementioned DSL which describes the system's context; for the running example this could be for example an intersection in an urban-like environment with some trees, buildings, and of course moving traffic. The last application which is called "monitor" is used to inspect non-reactively a running system or even a system simulation without modification. Using this application, the data at any stage of the processing chain can be visualized in various representations: Embedded into a 3D context, aggregated in charts, or any desired representation by easily extending the plug-in-based monitor application.

### 3.2.2 Sensor Raw Data Provider

As described in the previous section, the "Virtualization Layer" is used to generate different input data for the aforementioned layers. While the required input data for the "Action Layer" is rather simple to describe and thus to generate necessary input data for, the "Decision Layer" requires an abstract representation of the perceived system's surroundings. However, this representation can be modeled and thus provided for this layer with manageable effort. For the left-most layer which is dealing with the gathering and processing of the sensors' raw data, the model and generation of the required input data is rather complex.

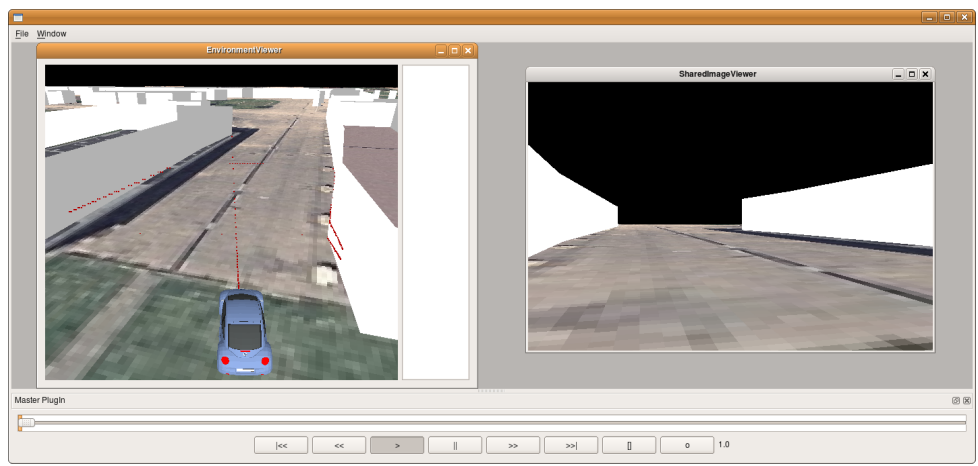

Figure 3: Application "monitor" which visualizes synthetic sensors' raw data: On the left hand side three single-layer laser scanners are modeled which are used to detect obstacles from the vehicle's surroundings; on the right hand side input data from a color camera is depicted. Both data is generated from one single source following the single point of truth *(SPOT)* principle: The DSL of the system's context.

In Fig. 3, the application "monitor" is shown which visualizes non-reactively synthetic sensors' raw data from three independent single-layer laser scanners and one color camera.

For generating the input data for the laser scanners, a GPU-based algorithm is implemented as outlined in [22]. Therefore, the DSL which describes the system's context is transformed automatically into a render-able representation by OpenGL at run-time. This render-able representation is loaded on the GPU and analyzed using the aforementioned algorithm. The result is transformed into the sensor specific data format and sent to all interested applications. The same OpenGL representation of the DSL can be used to provide images to simulate a color camera as well.

# 4  Application: Hella IDIS

The Hella IDIS LIDAR sensor is marketed as a cost-effective infrared distance sensor. It provides the lateral position and width of objects and can be mounted into the radiator grille of a car. It has a range of 3-150 meters, and a field of view of 12 degrees (US model) horizontally and 3 degrees vertically. While Hella makes a model that allows the sensor to estimate the trajectories of objects, if it is provided with additional data from the car, we opted for raw data output and fusing the data ourselves.

## 4.1  Technical Integration

Our IDIS was delivered with mounting brackets, connectors, and instructions for standard car installation. For on-car use, a weather-proof US-car connector is needed; however, for bench testing a simple molex connector will do. We install the sensor on our car's bull bar for evaluation; we will ultimately mount it between the radiator and the plastic grille.

## 4.2  Software Integration

Integrating the Hella IDIS into IRT requires two steps: *(i)* create a chardevice driver for a CAN interface, as this is our first CAN-connected sensor, and *(ii)* create the sensor driver itself.

The implementation of CanCharDevice is system-specific; we handle this using conditional compilation. Although in practice we will use the NetBurner version of CanCharDevice with the IDIS, for simplicity we only present here the version for SocketCAN [23] on Linux. SocketCAN was originally developed by Volkswagen Research under the name "Low Level CAN Framework" (LLCF); it extends the Berkeley sockets API with a new protocol family, PF_CAN, and is supported by the Linux kernel in version 2.6.25 and above. Our CAN-to-USB adaptors are SocketCAN-compatible. Listing 1 shows a simple implementation of CanCharDevice for SocketCAN / Linux with only basic error handling.

Listing 1: A (simplified and edited for space) implementation of CanCharDevice for Socket-CAN / Linux

```cpp
#include <chardevice/chardevice.h>
#include <can/can.h>

// Implementation file portion of registering CanCharDevice as
// a CharDevice plugin.
PLUGIN_DEFINE(CanCharDevice)

int CanCharDevice::open(void) {
    struct sockaddr_can addr;
    struct ifreq ifr;
    char *tokens[2];
    int numTokens = 0;
    s = ::socket(PF_CAN, SOCK_RAW, CAN_RAW);
    if(s < 0) exit(1);
    strcpy(ifr.ifr_name, "can0"); // default to can0
    if(strlen(options) > 0) {
        numTokens = str_split_destroy(options, ',', tokens, 2);
        if((numTokens == 1) && (tokens[0] != NULL))
            strcpy(ifr.ifr_name, tokens[0]);
    }
    ioctl(s, SIOCGIFINDEX, &ifr);
    addr.can_family = AF_CAN;
    addr.can_ifindex = ifr.ifr_ifindex;
    bind(s, (struct sockaddr *)&addr, sizeof(addr));
    return 1;
}

ssize_t CanCharDevice::read(void *buf, size_t count) {
    if(count < sizeof(struct can_frame)) return -1;
    else {
        struct ::can_frame frame;
        int nbytes = ::read(s, &frame, sizeof(frame));
        if(nbytes != 16 || frame.len + 2 > count) return -1;
        *((uint16_t *)buf) = frame.id;
        memcpy(&(((uint8_t *)buf)[2]), frame.data, frame.len);
        return frame.len + 2;
    }
}

int CanCharDevice::close(void) {
    return ::close(s);
}
```

Now that we have a chardevice implementation that supports CAN, we can test it using a GUI that is already built into IRT: `cdterm "can,can0"` ("can" specifies the CAN plugin, and "can0" is the configuration given to the CAN plugin—it specifies the name of the CAN interface). In the GUI, we can connect and disconnect; when we are connected, we will see all of the CAN messages on the bus. If we implemented CanCharDevice::write(), we could use the cdterm GUI to write messages to the CAN bus.

The next step is to write the IDIS sensor driver. Since we cannot yet release the exact interface of the IDIS, we will only outline some of the steps in this section. However, we hope to be able to release the complete source code soon, including this driver.

Although the IDIS is a LIDAR, it is conceptually more similar to a radar in that it returns information about a small set of obstacles instead of a dense set of range data. Thus, all that is left to do is to write an plugin for the radar driver subsystem; the only functions that we need to implement are for initializing the sensor, and for reading a dataset.

We can read CAN data from chardevice like this:

Listing 2: Reading CAN data using chardevice

```cpp
#include <chardevice/chardevice.h>

uint8_t buf[18];
// false means do not wait for data when call read()
CharDevice *cd = new CharDeviceWrapper("can,can0", false);
// read any available CAN message
int n = cd->read(buf, 18);
if (n > 2) {
    // print message info (first 2 bytes are message ID)
    printf("id = %hx (%d bytes)\n", *((uint16_t*)buf), n - 2);
}
// when finished, close chardevice
delete cd;
```

In HellaRadar::init(), we open the chardevice. In HellaRadar::read() we call the non-blocking chardevice read() method until no CAN message is returned; if a returned CAN message has an ID of interest, we extract the relevent information and continue reading CAN messages—we always return the most recent information available from the sensor.

Now that we have a Hella plugin for the radar subsystem, we can use it like any other radar driver—we can, for example, run the driver in server mode on a NetBurner board, which will timestamp the data (using a clock maintained from its external trigger signal) and forward it over Ethernet to any vehicle computer that is running a radar driver in client mode that has connected to that NetBurner's radar server (using e.g. the Spread Toolkit).

## 4.3 Calibration

In order to effectively fuse data from a camera, a Hella IDIS, and a SICK LMS laser scanner, we must determine the rigid body transformation between each pair of sensors. We choose the focal point of the camera as the origin of the local coordinate system; we must then determine the extrinsic calibration of the two laser scanners with respect to the camera. Once we have determined these transformations, the 3D coordinates from each laser scanner can be transformed to the camera's coordinates and back-projected onto the camera images for further processing.

In order to determine the 3D vector (through the camera focal point) on which the point represented by a given pixel is constrained to lie, we must obtain the intrinsic calibration of the camera. We determine these internal parameters of the camera by using the Caltech camera calibration toolbox [24].

We compute the extrinsic calibration between the camera and a given laser scanner only once, as the sensors are rigidly mounted relative to each other. We perform this calibration using at least three points in 3D space that we can identify in both sensors—the laser scanner (either the IDIS or the SICK) measures the 3D location of the point in its coordinate system directly, and we can convert the vector through the camera pixel to which the point is projected into a 3D coordinate in the camera's coordinate system by using space resection. Within a RANSAC [25] framework, we use the 3-point algorithm [26] to determine the depth of the points in the camera coordinate system and then use Horn's method [27] to find the rigid transformation between the two sets of 3D point correspondences. The geometry is illustrated in Figure 4, and the procedure is explained in full detail in [28, 29].

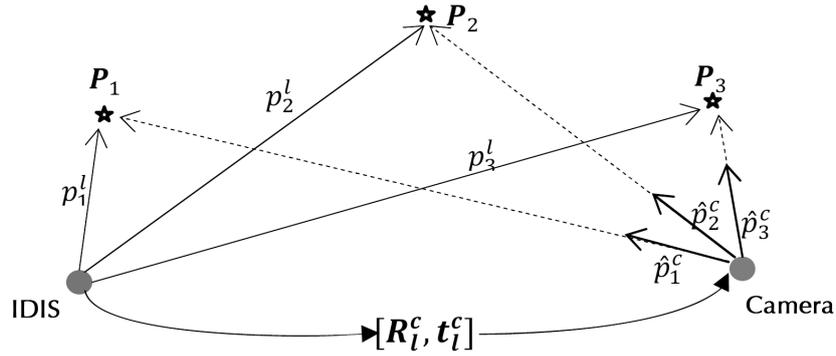

Figure 4: Three pairs of correspondences between the camera and the laser scanner serve to determine the relative sensor pose.

# 5   Conclusion & Outlook

In this paper we have introduced the Berkeley DRIVE Lab: our infrastructure, our hardware architecture, and our software toolkit BART, a fusion of Berkeley's Intelligent Robotics Toolkit (IRT) and Aachen's Hesperia Software Environment. We have also showcased our software's usability by demonstrating the rapid integration and calibration of a new automotive sensor. We hope to soon finish a final clean-up of the software and its documentation, and release it to the general public under a three-clause BSD license.

We would like to thank everyone who chatted with us during a DARPA race, showcased their vehicles to us, gave talks to us, and provided information about their vehicle hardware and software architectures online.

We hope that our work inspires others, and we invite both feedback and participation. Thank you.

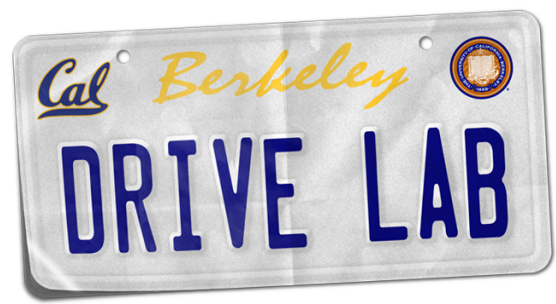